\documentclass[aps,twocolumn,pra,amsmath,amssymb,amsfonts,superscriptaddress]{revtex4-1} 
\usepackage[T1]{fontenc}
\usepackage{stix}
\usepackage[utf8]{inputenc}
\usepackage{verbatim}
\usepackage{textcomp}
\usepackage{amsmath}
\usepackage{amssymb}
\usepackage{graphicx}
\usepackage{color}
\usepackage[english]{babel}
\usepackage{booktabs}
\usepackage{multirow}
\usepackage{tabularx}
\begin{document}
\title{Quantum metrology with Bloch Oscillations in Floquet phase space}
\author{Keye Zhang}
\email{kyzhang@phy.ecnu.edu.cn}
\affiliation{Quantum Institute for Light and Atoms, State Key Laboratory of Precision Spectroscopy, School of Physics and Electronic Science, East China Normal University, Shanghai 200241, China}
\affiliation{Shanghai Branch, Hefei National Laboratory, Shanghai 201315, China}
\author{Weijie Liang}
\affiliation{Quantum Institute for Light and Atoms, State Key Laboratory of Precision Spectroscopy, School of Physics and Electronic Science, East China Normal University, Shanghai 200241, China}

\author{Pierre Meystre}
\affiliation{Department of Physics and College of Optical Sciences, University of Arizona, Tucson, AZ 85721, USA.}

\author{Weiping Zhang}
\email{wpz@sjtu.edu.cn}
\affiliation{School of Physics and Astronomy, and Tsung-Dao Lee Institute, Shanghai Jiao Tong University, Shanghai 200240, China}
\affiliation{Shanghai Research Center for Quantum Sciences, Shanghai 201315, China}
\affiliation{Shanghai Branch, Hefei National Laboratory, Shanghai 201315, China}
\affiliation{Collaborative Innovation Center of Extreme Optics, Shanxi University, Taiyuan, Shanxi 030006, China}


\begin{abstract}
Quantum particles performing Bloch oscillations in a spatially periodic potential can be used as a very accurate detector of constant forces. We find that the similar oscillations that can appear in the Floquet phase space of a quantum particle subjected to a periodic temporal driving, even in the absence of periodic lattice potential, can likewise be exploited as detectors. Compared with their spatial Bloch analog, however, the Floquet-Bloch oscillations provide significant added flexibility and open the way to a broad range of precision metrology applications. We illustrate this property with the examples of a tachometer and a magnetometer.  
\end{abstract}

\maketitle
\section{Introduction}
It is well known that quantum particles confined in periodic potentials and subjected to a constant force do not accelerate uniformly in real space. Rather, they undergo Bloch oscillations, which have been observed in a variety of systems, including ultracold atom gases trapped on optical lattices~\cite{Dahan1996, Wilkinson1996, Gustavsson2008} and optical waves in waveguide arrays~\cite{Pertsch1999, Morandotti1999}. Since the Bloch oscillations frequency depends only on the lattice spacing and the applied external force it provides an accurate tool for the measurement of weak forces and the precise determination of fundamental constants~\cite{Poli2011, Biraben2004, Biraben2006, Biraben2011}.

There is much interest in extending the study of these systems to situations where they are subjected in addition to a time-periodic forcing~\cite{Arlinghaus2011, Kudo2011}. This includes in particular the study of `super-Bloch oscillations', which can result in linear transport in a lattice~\cite{Haller2010} and spatial oscillations of the system even in the absence of a spatial lattice~\cite{Junk2020}. More generally, time-periodic forcing is now exploited in the emerging area of `Floquet engineering', a powerful tool to control and modify quantum systems, including for instance the discovery of new out-of-equilibrium phases, see e.g.~\cite{Moessner2017, Weitenberg2021}, in particular, the Floquet time-crystal phase~\cite{Else2016, Sacha2018a}. Applications of these techniques to enhance the precision of quantum metrology have also been proposed, see e.g.~\cite{Lukin2020, Fiderer2018, Jiang2021}.

The present work extends the Bloch measurement idea to the Bloch-like oscillations that appear in the Floquet phase space of a system subjected to a periodic temporal drive rather than a spatially periodic potential. Just like in the case of Bloch oscillations, the measurement of constant forces applied to systems undergoing such oscillations reduces to frequency measurements, with the remarkable precision with which they can be carried out. But as we shall show, the transformation of the Hamiltonian of the `measuring apparatus' to the phase representation brings to the fore two key features of this approach, as compared to the familiar Bloch oscillations approach: It is not limited to the measurement of constant forces, and it applies to a variety of physical observables. We will illustrate these points on the specific examples of the realization of a tachometer and a magnetometer. 

The paper is organized as follows: Section II presents the analytical analysis of the underlying idea under a secular approximation, which is then revisited via numerical simulations in Section III. Section IV then applies our proposed scheme approach to the examples of a tachometer and a magnetometer. Finally, Section V is a summary and outlook.  Some additional details of the relevant action-angle transformations are given in an Appendix. 

\section{Bloch Dynamics in Floquet phase space}
We consider a system described by the one-dimensional Hamiltonian 
\begin{equation}
    H = H_0 + H_1 = \frac{p^2}{2m} +U(x) + V(x)\cos\omega t \,,
	\label{H1D}
\end{equation}
where $U(x)$ is a conservative and non-periodic potential and $H_1= V(x)\cos\omega t$ is a weak single-frequency perturbation oscillating at the frequency $\omega$. 

Two common strategies can be used to reveal the lattice structure of this system in Floquet phase space. The first one is based on a perturbative approach combined with a specific time-dependent canonical transformation~\cite{Berman1977, Berman1981, Berman1985, Buchleitner2002}, while the second one relies on the use of Floquet quantum states and quasi-energy theory~\cite{Grifoni1998, Holthaus2016}. We adopt the first method here, as it is particularly convenient to discuss the various measurement applications that we have in mind.

\subsection{Lattice structure in Floquet space}

Consider first the classical version of the unperturbed Hamiltonian $H_0$. For a given system energy $H_0 = E_0$ it is always possible to introduce a pair of conjugate action-angle variables $(J_{0}, \Theta)$ with 
\begin{equation}
J_0 = \frac{1}{2\pi}\oint p \,{\rm d} x =  \frac{1}{2\pi}\oint \sqrt{2m[E_0-U(x)]} {\rm d} x\, ,
\label{eq:J}
\end{equation}
where the integral is over one period of motion, so that the transformed Hamiltonian $H_0$ depends on the action variable $J_{0}$ only. The associated angle variable $\Theta$ evolves at a constant angular frequency $\Omega$ in the range $\left[0,2\pi\right)$, with Hamilton equations of motion
\begin{equation}
\dot{J_0}=-\frac{\partial H_0(J_0)}{\partial\Theta}=0\,\,\,\,;\,\,\,\dot{\Theta}=\frac{\partial H_0(J_0)}{\partial J_0}=\Omega\, .
\end{equation}

In terms of these variables the perturbation potential $V(x)\cos\omega t$ is resolved into multiple components as a Fourier series in $\Theta$,
\begin{equation}
V(x) \to V(J_{0} ,\Theta)=\sum_{\ell=-\infty}^{+\infty}V_{\ell}(J_{0})e^{{\rm i}\ell \Theta} \, ,
\end{equation}
where $V_\ell(J_{0})$ are the Fourier components of $V(x)$ evaluated along the unperturbed classical trajectory of energy $E_0$,
\begin{equation}
	V_{\ell}(J_0)=\frac{1}{2\pi}\int_{0}^{2\pi}V(x)e^{-i\ell\Theta}d\Theta\, .
	\label{Vell}
\end{equation}

For small perturbations the phase space trajectories of the perturbed dynamics will remain close to the unperturbed ones, that is $\Theta\approx\Omega t$ and $\ell\Theta-\omega t\approx(\ell\Omega-\omega)t$.  If the frequency $\omega$ of the perturbation is a multiple of the unperturbed angular frequency,
\begin{equation}
	\omega\approx n\Omega,\:n=1,2,3\cdots,
\end{equation}
all components in that sum will oscillate rapidly and average out to zero, except for the near-resonant term $V_{n}\cos[n(\Theta-\Omega t)]$. It would therefore seem reasonable to discard the fast-oscillating terms in a familiar secular approximation~\cite{Zhang2017}. We proceed along these lines in the following analytical approach but will return to this point in the next section to estimate via a full numerical simulation the limitations of this approach.

Following Ref.~\cite{Buchleitner2002} we then carry out a second time-dependent canonical coordinate transformation involving the slowly varying variable 
\begin{align}
	\vartheta & =\Theta-\frac{\omega t}{n}\, ,
\end{align}
so that the effective secular Hamiltonian becomes a time-independent one in
the new phase space $(J,\vartheta)$,
\begin{equation}
	H_{0}+H_{1}\approx H_{0}(J)+V_{n}(J)\cos n\vartheta-\frac{\omega J}{n}\, ,\label{eq:Heff1}
\end{equation}
where the last term, $-\omega J/n$, is the contribution of the canonical momentum conjugate to time~\cite{footnote}. Because this Hamiltonian is nothing but the classical analog of the quantum Floquet Hamiltonian we call the extended phase space $(J,\vartheta)$ the Floquet phase space in the following.

For small $V_{n}$ we perform a power expansion of the unperturbed Hamiltonian $H_0+H_1$ about the resonant action $J_{0}$,
\begin{widetext}
	\begin{equation}
		H_{0}+H_{1}\approx H_{0}(J_{0})-\frac{\omega J_{0}}{n}+\left(\left.\frac{\partial H_{0}}{\partial J}\right|_{J=J_{0}}-\frac{\omega}{n}\right)(J-J_{0})+\frac{1}{2}\left.\frac{\partial^{2}H_{0}}{\partial J^{2}}\right|_{J=J_{0}}(J-J_{0})^{2}+V_{n}(J_{0})\cos n\vartheta\, ,
	\end{equation}
\end{widetext}
where the constant zeroth-order term can be removed and the first-order term is zero due to the fact that $J_0$ is the action corresponding to the resonant condition $\omega=n\Omega$. Introducing a new `conjugate momentum'
\begin{equation}
	P=J-J_{0} \, ,
\end{equation}
and the `effective mass'
\begin{equation}
	M=(\partial^{2}H_0/\partial J^{2}_0)^{-1} \, ,
    \label{Meff}
\end{equation}
it follows that $H_0+H_1$ reduces to the Hamiltonian of a particle in a periodic lattice potential,
\begin{equation}
	H_{0}+H_{1}\approx\frac{P^{2}}{2M}+V_{n}(J_{0})\cos n\vartheta \, .
    \label{eq:Heff2}
\end{equation}
Notice that the units of $P$ and $M$ are the same as angular momentum and moment of inertia respectively since the Floquet phase space is based on the angle variable.

The transforms from $(x,p)$ to $(J_0,\Theta)$ and to $(P,\vartheta)$ are all canonical since their Poisson brackets are unity. However, in the quantum counterpart of the action-angle transform, the fact that $P$ now has a discrete spectrum implies that the canonical commutation relation $[\vartheta, P]=i\hbar$ only holds if the wave function $\varphi(\vartheta)$ is periodic with period $2\pi$ \cite{Rabitz1979, Harris1996}. When quantizing $H(P, \vartheta)$ by promoting $P$ to an operator, $P\rightarrow-{\rm i} \hbar\partial_\vartheta$, its eigenfunctions will therefore be Bloch-like functions of the form
\begin{equation}
    \varphi_{q}(\vartheta)=U_{q}(\vartheta)e^{{\rm i}q\vartheta} \, ,
\end{equation}
where $U_q(\vartheta)$ has the periodicity $2\pi/n$ of the potential $V_n \cos n\vartheta$, but with the caveat that the effective `quasimomenta' $q$ have to be dimensionless integers to meet the periodicity requirement of the wave functions. Within the secular approximation, these implicitly time-dependent Bloch-like functions are the Floquet eigenstates~\cite{Sacha2018}.

In the first Brillouin zone, the `quasi-momenta' $q$ are integers in the range $[0,n)$, with dispersion relations $\varepsilon_{m,q}$, so that a general state of the system takes the form
\begin{equation}
\psi(\vartheta,t)=\sum_{m=1}^{\infty}\sum_{q=0}^{n-1}c_{m,q}(0)\varphi_{m,q}(\vartheta)e^{-{\rm i}\varepsilon_{m,q} t/\hbar}\, .
\label{eq:psit}
\end{equation}
Here $m$ is the band index and $c_{m,q}(0) = \int \psi(\vartheta,0) \varphi_{m,q}^* (\vartheta)\,{\rm d}\vartheta$ are the initial probability amplitudes of the Bloch-like functions $\varphi_{m,q}$. 

\subsection{Measurements by Floquet-Bloch oscillations}

We now discuss how systems undergoing such Floquet-Bloch oscillations can be exploited as a `measuring apparatus' that allows under a broad range of conditions for the determination of forces via frequency measurements. We focus for simplicity on the example of a weak force $f$ linearly coupled to $\vartheta$ in Floquet space via the interaction
\begin{equation}
V_{\rm probe} = f \vartheta \, .
\label{eq:vprobe} 
\end{equation}
As we shall see shortly, this does not imply that we limit ourselves to linear interactions in physical space.  As a result, the `quasimomentum' $q$ undergoes the evolution
\begin{equation}
q(t)=q(0)-ft/\hbar \, .
\end{equation}
Assuming that the system initially populates only the first band and that $V_{\rm probe}$ is weak enough that it does not induce interband transitions the state~(\ref{eq:psit}) becomes approximaly~\cite{Gluck2002}
\begin{equation}
\psi(\vartheta,t)\approx \sum_{q=0}^{n-1}c_q(0) \varphi_{q(t)}(\vartheta) e^{-\frac{\rm i}{\hbar}\int_{0}^{t}\varepsilon_{q(t)}{\rm d} t} \, ,
\end{equation}
where we ignore the band index $m=1$ for notational convenience. The group velocity of this wavepacket is $v_g\sim\partial\varepsilon(q)/\hbar\partial q$, and its `effective mass' $m_g\sim[\partial^2\varepsilon(q)/\partial q^2]^{-1}$, oscillating between positive and negative values. As a result, the wave packet moves back and forth along $\vartheta$ in Floquet phase space with the Bloch period
\begin{equation}
T_{\rm B} = \frac{\hbar n}{|f|}\, , 
\label{eq:TB}
\end{equation}
the time it takes $q(t)$ to propagate from $q(0)$ to $q(0) + n$ across the  Brillouin zone.

The amplitude of the Floquet-Bloch oscillations is ${\cal W}/2f$, where $\cal W$ is the bandwidth of the first band \cite{Hartmann2004}. For large normalized lattice depths $s=V_n/E_n \gg 1$, with $E_n\equiv \hbar^2n^2/2M$, that bandwidth can be estimated as~\cite{Bloch2008}
\begin{equation}
    {\cal W}\approx\frac{4}{\sqrt{\pi}}(8s)^{3/4}e^{-4\sqrt{2s}} E_n,
	\label{bandwidth}
\end{equation}
indicating that the amplitude of the oscillations is exponentially suppressed for increasing lattice depth.
A more precise estimate that holds for shallower lattices can be obtained by invoking a Wentzel–Kramers–Brillouin approximation~\cite{Guo2013}.

Returning now to the original $(P, \Theta)$ picture, we have that the system oscillates in addition at the unperturbed frequency $\Omega$. With $\vartheta = \Theta - \Omega t$ and $\omega = n \Omega$ we find that $\Theta(t+T_{\rm B}) - \Theta(t) =  \Omega T_{\rm B} = \hbar \omega/|f|$. Therefore, after each Bloch period $\psi(\Theta,t)$ accumulates a drift $\hbar \omega/f$ in the angle variable $\Theta$,  so that
\begin{equation}
\psi(\Theta,t)=\psi \big  (\Theta+\frac{\hbar\omega}{f},t+T_{\rm B} \big ) = \psi \big  (\Theta+\frac{\hbar\omega}{f},t+\frac{\hbar n}{f}\big )\, .
\label{psiperiod}
\end{equation}
which shows that it is possible to exploit either the period of oscillations or the phase shift of $\psi(\Theta,t)$ to determine $f$. This is the central result of this paper. 

\subsection{Implementation} 
One simple way to realize potentials of the form~(\ref{eq:vprobe}) is by using a train of $\delta$-kicks separated by an interval $T_{\rm D}=2\pi/\Omega$. This follows from the fact that we have then
\begin{eqnarray}
V_{\rm probe} &=&f\Theta \sum_{\ell=-\infty}^{\infty}T_{\rm D}\, \delta(t-lT_{\rm D}) \nonumber \\
&=& f\left [\pi -\sum_{ k =1}^\infty \frac{2 \sin k\Theta}{k} \right ] \sum_{\ell = -\infty}^\infty e^{{\rm i} \ell \Omega t} 
\approx f \vartheta \, ,\
\label{secapp}
\end{eqnarray}
where the first summation in the second line is the Fourier series of  $\Theta$ for $0< \Theta <2\pi$, the second summation is the Fourier series of the Dirac comb function, and the final approximate equality holds under the secular approximation. 

Since the relationship between the phase coordinates $\Theta$ and the physical coordinates $(x,p)$ depends on the explicit form of $H_{0}$ it follows that the Hamiltonian $V_{\rm probe}$ can describe many physical processes, in contrast with the situation with Bloch oscillations in a spatial lattice. Floquet-Bloch oscillations in phase lattice can therefore be exploited for the measurement of a broader variety of observables by an appropriate design of the conservative potential $U(x)$ in $H_0$. 

\section{Numerical Simulations}

Before illustrating this measurement technique on several concrete examples we first need to revisit the assumptions and limitations underlying Eq.~(\ref{psiperiod}), as they inform the range of experimental situations achievable in practice. In particular, we have seen that the amplitude of the Bloch oscillations is exponentially suppressed by the lattice depth. This feature needs to be reconciled with our assumption that only the first band of the system is initially populated, which requires that the band gap, roughly equal to $V_n$, be much larger than $V_{\rm probe}$, which scales as $(2\pi/n) f$.  In addition, the secular approximation holds only under the perturbation condition $V_n \ll \hbar\Omega$, since otherwise the off-resonant driving becomes significant and can result in irregular dynamics.

To address these issues more quantitatively we have carried out detailed numerical simulations for the case of a particle of mass $m$ trapped in an infinite square well $U(x)$ of width $L$ and subjected to a single-frequency periodic driving at frequency $\omega$ and a sequence of $\delta$-kicks in alternating directions, with Hamiltonian
\begin{equation}
H= \frac{p^2}{2m}+U(x)+ gx\cos \omega t + {\rm sign}(p) F_0 \sum_{l=-\infty}^{+\infty} \delta(t-l T_{\rm D}) x \, .
\label{Hsquare}
\end{equation}
For this square potential, the action-angle transformation between the position $x$ and $\Theta$ is linear, 
\begin{equation}
x=L\left|\pi-\Theta\right|/\pi \, ,
\label{eq:x}
\end{equation}
and the angular frequency is $\Omega=\sqrt{2\pi^2 E_0/mL^2}$ with $E_0$ the initial energy, see the Appendix for the details of the transformation.
For a Gaussian wave packet with a phase factor $e^{ip_0x/\hbar}$, which corresponds to a classical particle with initial kinetic energy $E_0=p_0^2/2m$, the angular frequency becomes $\Omega=\pi p_0/mL$, the oscillation frequency of a wave packet in a square well of width $L$. 

As derived in the previous section, in the Floquet phase space the periodic driving is resolved into multiple components as a Fourier series in the angle $\Theta$, 
\begin{equation}
	gx\cos \omega t=\sum_{\ell=-\infty}^{+\infty}V_{\ell}e^{{\rm i}\ell \Theta}	\cos n\Omega t\, ,
\end{equation}
so that, with the integral (\ref{Vell}) and the transform (\ref{eq:x}), the only non-vanishing amplitudes are  for $\ell$ odd, $V_{\ell}=2gL/\pi^2\ell^2$. Under the secular approximation only the components $\ell=\pm n$ survive, resulting in an effective $n$-site lattice potential, $(2gL/\pi^2 n^2)\cos n\vartheta$.

Similarly, in the Floquet phase space, the sequence of $\delta$-kicks is resolved into multiple components
\begin{equation}
	{\rm sign}(p) F_0 \sum_{l=-\infty}^{+\infty} \delta(t-l T_{\rm D}) x = \frac{LF_0}{\pi}\sum_{k=1}^\infty\frac{2\sin k\Theta}{k}\sum_{l=-\infty}^{\infty}\frac{\tau}{T_{\rm D}}e^{il\Omega t},
\end{equation}
where the kick duration $\tau \ll T_{\rm D}$. The argument leading to Eq.~(\ref{secapp}) shows that under the secular approximation, they act effectively as the linear potential $f\vartheta$ due to a constant weak force $f=LF_0\tau/\pi T_{\rm D}$, resulting in the Floquet-Bloch oscillations with the period $T_{\rm B}=\hbar n\pi T_{\rm D}/LF_0\tau$. Then the amplitude of the kick force can be determined by direct measurement of the frequency of the Floquet-Bloch oscillations as
\begin{equation}
    F_0=\frac{\hbar n\pi}{\tau} \sqrt{\frac{2m}{E_0}}\left ( \frac{1}{T_{\rm B}} \right ) \, .
\end{equation}

In our simulation, we used the original Hamiltonian (\ref{Hsquare}) in coordinate space instead of the perturbative Floquet space lattice Hamiltonian obtained in the secular approximation. We also accounted for the decoherence induced by thermal motion by describing the evolution of the quantum state of the particle via the master equation with quantum Brownian noise~\cite{Gardiner1999, Caldeira1983}
\begin{equation}
	\frac{d\rho}{dt}-=\frac{i}{\hbar}[H,\rho]-\frac{i\gamma}{\hbar}[\hat x,[\hat p,\rho]_+]-\frac{2m\gamma}{\hbar^2\beta}[\hat x,[\hat x,\rho]]\, ,
\end{equation}
with $\gamma$ and $\beta$ the damping rate and the inverse temperature, respectively.

\begin{figure}
	\includegraphics[width=3.4in]{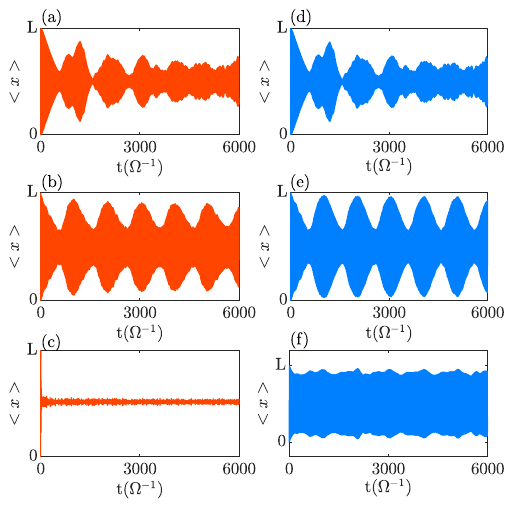}
	\caption{Evolution of the expectation value $\langle x(t)\rangle$ of the position of a quantum particle trapped in an infinite square well and subjected to both a periodic driving with frequency ratio $n\equiv \Omega/\omega= 31$ and a sequence of $\delta$-kicks at intervals $T_{\rm D}= 2\pi/\Omega$, in the absence of dissipation. Here the effective force is $f = F_0 L \tau / \pi T_{\rm D} = 1.6\times 10^{-3}\hbar\Omega/2\pi$, so that the scale of $V_{\rm probe}$ is $(2\pi/n)f=5\times10^{-5}\hbar\Omega$, and the time is in units of $1/\Omega$. The left red curves show the exact evolution, without the secular approximation and the right blue curves the corresponding evolution under the secular approximation.Top: $V_n=0.0009\hbar\Omega$; Middle: $V_n=0.09\hbar\Omega$; Bottom: $V_n=0.9\hbar\Omega$. }
		 
\label{oscillation1}
\end{figure}

The results of the simulations with a fixed value of $n=\Omega/\omega$ and for several values $V_n$ are summarized in Fig.~\ref{oscillation1}, which compares the time evolution of the mean position $\langle x(t) \rangle$ without (left red plots) and with (right blue plots) the secular approximation. For $V_n\ll \hbar\Omega$, the two solutions almost overlap, confirming the validity of the secular approximation in that regime.  However, larger values of $V_n$ result in an increased discrepancy between the exact and approximate results, confirming that the approximation only holds provided that $H_0\gg H_1$. In particular, for $V_n$ close to $\hbar\Omega$, $\langle \hat x(t)\rangle$ exhibits a disordered evolution completely absent from the secular approximation behavior. 

Fig.~\ref{oscillation1} also illustrates that the Floquet-Bloch oscillations evident in the periodic changes of the envelope of  $\langle x(t)\rangle$  are only visible for intermediate values of $V_n$. More specifically, in the top figure ($V_n = 0.0009 \hbar \Omega$) we have that $V_n \lesssim V_{\rm probe}$ and the assumption underlying our analysis that $V_{\rm probe}$ does not induce interband transitions ceases to hold, even though the secular approximation works well. This shows that the proposed measurement scheme requires that $V_n$ satisfies the `Goldilocks condition' of being large enough that no interband transition takes place, but small enough that the secular approximation holds approximatively. 

\begin{figure}
	\includegraphics[width=3.4in]{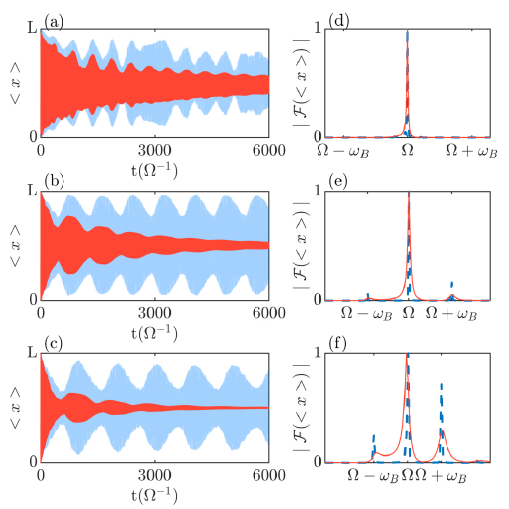}
	\caption{Evolution of the mean position $\langle x(t)\rangle$ and its Fourier spectrum $|{\cal F}(\langle x\rangle)|$ of a quantum particle in an infinite square well subjected to a periodic driving for $V_n=0.09\hbar\Omega$ and $\delta$-kicks resulting in an effective force $f=1.6\times 10^{-3}\hbar\Omega/2\pi$ and subject to quantum Brownian noise with $\gamma=1.34\times 10^{-6}\Omega$, for several values of the frequency ratio $n=\Omega/\omega$. Left: time evolution of $\langle\hat x(t)\rangle$ (dark red curves); Right: its Fourier spectrum (red solid curves), Top: $n=15$; Middle: $n=25$; Bottom: $n=31$. The right blue (light) curves and the left blue dashed curves are for the cases in the absence of dissipation.}
		 
	\label{oscillation2}
	\end{figure}

The effects of dissipation are shown in Fig.~\ref{oscillation2}, which compares $\langle x(t)\rangle$ in the presence of dissipation resulting from coupling to a Brownian reservoir to the case without dissipation. In these plots, $V_n$ is kept constant, but the frequency ratio $n=\Omega/\omega$ takes increasing values $n=$ 15, 21 and 31 from the upper to the lower plot, so as to illustrate both the dependence of the oscillation frequency on $n$ and the effects of the scaling of $V_{\rm probe}$ on $(2\pi/n) f$.

In the absence of dissipation, both the period and the amplitude of the Floquet-Bloch oscillations increase with increasing $n$, consistently with the expression (\ref{eq:TB}) for the oscillation period and with Eq.~(\ref{bandwidth}) which shows that the decrease of the normalized lattice depth $s$ for increasing $n$ results in an increase of the width of the first band, and hence an increase in the amplitude of oscillations. In Fourier space the amplitude of the Bloch frequency peaks at $\Omega\pm 2\pi/T_{\rm B}$ become therefore increasingly visible. Note also that for $n=15$, the top part of Fig.~\ref{oscillation2}, the oscillations in $\langle x(t)\rangle$ are strongly perturbed, a result of the appearance of interband transitions for the relatively large $V_{\rm probe}$ associated with this small $n$.

Turning now to the effect of dissipation, we observe that the damping of $\langle x(t)\rangle$ becomes faster as $n$ is increased. That is because the noise broadens the frequency spectrum of the system. For increased $n$ a correspondingly larger number of non-resonant low-frequency modes with $|\ell|<n$, which are ignored in the secular approximation, come into play. While they all have the same damping rate $\gamma$ for Brownian noise, their coupling to the resonant mode results in an increased decoherence rate $\kappa$ of the Floquet-Bloch oscillations. From our numerical simulations we estimate it scales approximately with $n$, $\kappa \sim n\gamma$. As a result, for increasing $n$ the frequency peaks first broaden and eventually disappear. 

Combined with the results summarized in Fig.~\ref{oscillation1}, these results illustrate that the frequency peaks of the Floquet-Bloch oscillations in the displacement spectrum are only distinct enough to determine $f$, and hence $F_0$, when the system parameters satisfy the inequalities
\begin{equation}
    \hbar\Omega \gg V_n \gg f2\pi/n \gg \hbar \kappa.
    \label{appcond}
\end{equation}
With these limitations in mind, we now turn to a secular approximation discussion of several potentials $U(x)$ appropriate for the Floquet-Bloch measurement of a variety of observables.

\section{Examples of measurement schemes}

We now turn a brief discussion of Floquet-Bloch measurement schemes based on other potentials $U(x)$ and demonstrate how they can be exploited, for instance, as tachometers or magnetometers. 

\begin{figure}
    \includegraphics[width=2.5in]{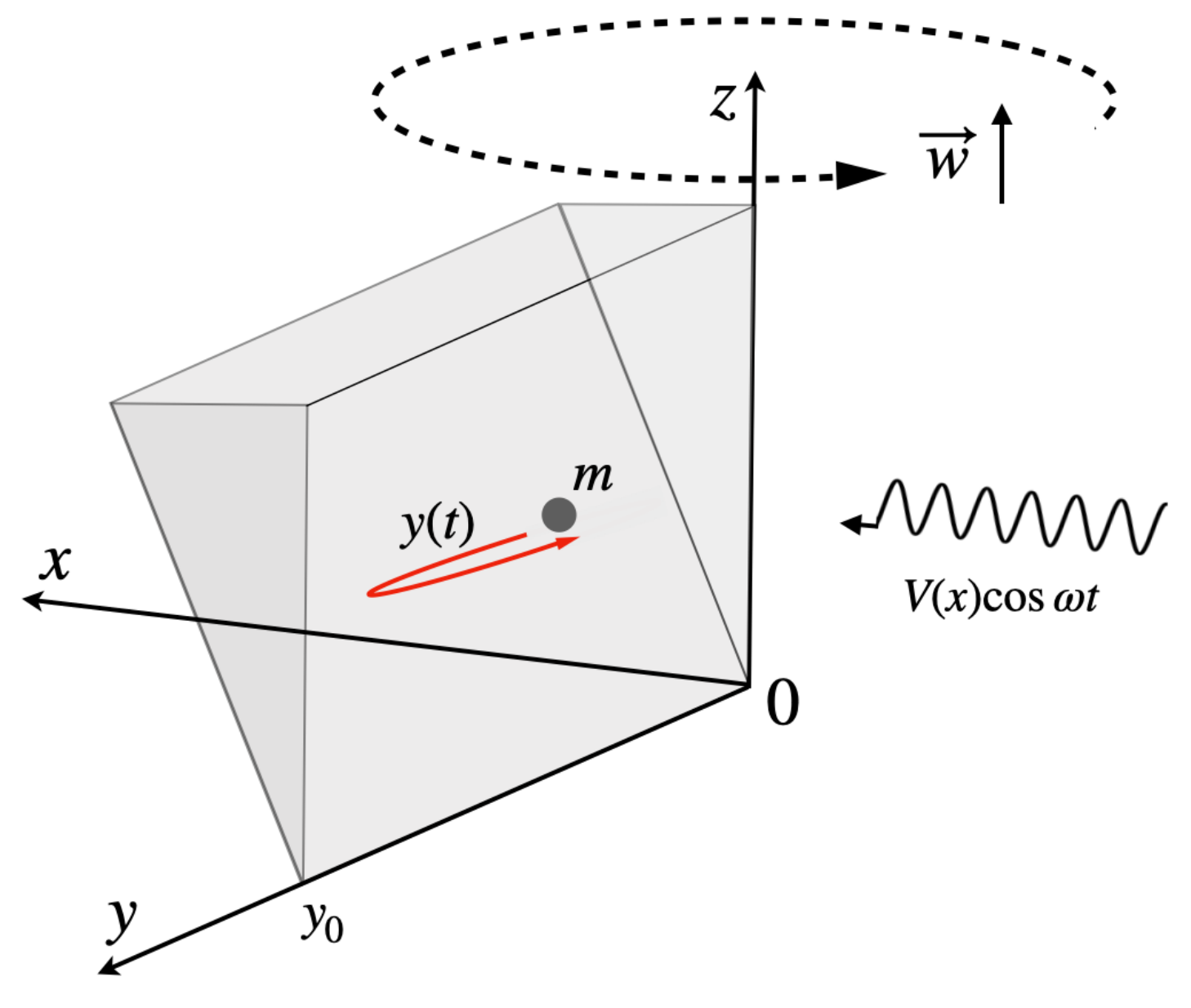}
    \caption{Tachometer based on Floquet-Bloch oscillations.
    }
    \label{tachometer}
\end{figure}

We consider first the one-dimensional triangular well potential 
\begin{equation}
    U(x)=\begin{cases}
    \eta x &  {\rm for} \; x \ge 0\\
    \infty & {\rm for} \; x < 0\, ,
    \end{cases}
\end{equation}
in which case the position and momentum $x$ and $p$ are related to the action-angle variables by~\cite{Buchleitner2002} 
\begin{eqnarray}
    x&=&\eta(2\pi\Theta-\Theta^2)/2m\Omega^2\,,\label{xtheta2}\\
    p&=&\eta(\pi-\Theta)/\Omega \, ,
    \label{ptheta2}
\end{eqnarray}
where the angular frequency is $\Omega = \eta\pi / \sqrt{2 m E_0}$, see the Appendix. Importantly, it is now the momentum that is linear in $\Theta$, suggesting that this potential might find applications in situations where $V_{\rm probe}$ depends linearly on the momentum $p$ rather than $x$. We present two examples that exploit such a situation, the first one resulting in the realization of a tachometer as shown in Fig.~\ref{tachometer}, and the second in a magnetometer.

\subsection{Tachometer}

The Hamiltonian for a particle of mass $m$ in a rotating frame of angular velocity $\vec{w}$ is
\begin{equation}
	H(\vec{r})=\frac{1}{2m}(\vec{p}-m\vec{w}\times\vec{r})^{2}-\frac{m}{2}(\vec{w}\times\vec{r})^{2}\, .
	\label{Hrot}
\end{equation}
If $\vec{w}$ is along the z-axis, and in the presence of the triangular well $U(x)$ and of a small periodic Floquet perturbation $V(x) \cos\omega t$, the effective two-dimensional Hamiltonian for the particle is then
\begin{equation}
	H_{xy}=\frac{p_{x}^{2}+p_{y}^{2}}{2m}+U(x)+V(x)\cos\omega t + w_z y p_x - w_z x p_y\, .
	\label{Hrotxy}
\end{equation}

One way to realize the one-dimensional Floquet Hamiltonian with a probe potential is by exerting control over the $y$ position of the particle. 
Considering in the rotating frame $p_y=m\dot{y}+mw_zx$, $H_{xy}$ gives the dynamical equations of $x$,
\begin{eqnarray}
	\dot{x}&=&\frac{p_x}{m}+w_z y\,, \\
	\dot{p}_x&=&-\frac{dU}{dx}-\frac{dV}{dx}+mw_z\dot{y}+mw^2_zx\,,
\end{eqnarray}
which can derive a reduced Hamiltonian for $x$ dimension when $y(t)$ is arranged in time,
\begin{equation}
	H_{x}=\frac{p_{x}^{2}}{2m}+U(x)-\frac{mw_z^2x^2}{2}+V(x)\cos\omega t + w_z y p_x - w_z m x \dot{y}\, .
	\label{Hrotx}
\end{equation}
We neglect the component $-mw_z^2x^2/2$ of the potential in the following derivation by assuming that the triangular well $U(x)$ is tight enough.

In practice the particle’s $y$ position is fast adjusted from $0$ to $y_0$ and back in a short duration $\tau$ by optical or magnetic trapping fields. If this adjustment is repeated every interval of period $T_{\rm D}$, the position evolution $y(t)$ can be approximated as a periodic Dirac function as long as $\tau\ll T_{\rm D}$,
\begin{equation}
    y(t)\approx y_0\tau \sum_{l=-\infty}^{+\infty}\delta(t-l T_{{\rm D}})=\frac{y_0\tau}{T_{\rm D}} \sum_{l=-\infty}^{+\infty} e^{i l \Omega t}\, ,
\end{equation} 
and then the last two terms in $H_x$ can be interpreted as a probe potential $V_{\rm probe}$ in the Floquet phase space. To see that this is the case, we apply the transform (\ref{ptheta2}) and invoke the secular approximation to the first term $w_{z} y p_x$, giving
\begin{equation}
    w_{z} y p_x=\frac{w_z y_0\tau}{T_{\rm D}} \sum_{l=-\infty}^{+\infty} e^{i l \Omega t}\frac{\eta}{\Omega}\sum_{k=1}^{\infty}\frac{2\sin k\Theta}{k}\approx\frac{w_z y_0\tau\eta}{2\pi}(\pi-\vartheta)\,,
\end{equation}
and similarly for the second term $-w_zxm\dot{y}$, 
\begin{eqnarray}
    -w_z xm\dot{y}&=& -\frac{w_z y_0\tau}{T_{\rm D}}\sum_{l=-\infty}^{+\infty} i l \Omega e^{i l \Omega t}\frac{\eta}{2\Omega^2}(\frac{8\pi^2}{3}-\sum_{k=1}^{\infty}\frac{4\cos k\Theta}{k^2}) \nonumber\\
    &\approx&\frac{w_zy_0\tau\eta}{2\pi}(\pi-\vartheta) \, ,
\end{eqnarray}
yielding the total probe potential 
\begin{equation}
    V_{\rm probe}=w_z(yp_x-xm\dot{y})\approx f(\vartheta-\pi)\, ,
\end{equation}
with
\begin{equation}
f=-\frac{w_z\eta y_0\tau}{\pi}\,.
\end{equation}

As a result, the system operates as a tachometer, with the Bloch frequency providing a direct measure of the angular velocity $w_{z}$ via
\begin{equation}
 w_z=\frac{\hbar n}{2 \eta y_0 \tau}\left ( \frac{2\pi}{T_{\rm B}} \right ) \, .
 \label{wz}
\end{equation}

As a concrete example consider a tachometer aimed at measuring angular velocities of the order of those of the fastest demonstrated spinning objects, with $w_z\sim 10^9$rad/s~\cite{Reimann2018, Ahn2018, Ahn2020}. For atomic scale particles of mass $m \sim 10^{-27}$kg whose motion can be controlled at the nanoscale level, $y_0\sim 10^{-9}$m and $\tau\sim 10^{-9}$s, an applied perturbation frequency of $\omega\sim 100\Omega\sim 10^{10}$Hz, and $\eta\sim 10^{-17}$N,  chosen to be much larger than the centripetal force $mw_z^2y_0$ to ensure that $V_{\rm probe}$ is the weakest perturbation, we find from Eq.~(\ref{wz}) that the Floquet-Bloch period is $T_{\rm B}\sim 10^{-6}$s, which is much shorter than the typical spatial Bloch periods, of the order of $10^{-3}$s, demonstrated for instance in ultracold atoms~\cite{Dahan1996}. Such short periods present the considerable advantage of imposing modest demands on the quantum coherence time of the observed objects -- we recall that the period of usual Bloch oscillations of electrons on solid-state lattices is typically much longer than their quantum coherence time so that their demonstration has been limited so far to artificial superlattices \cite{Feldmann1992, Leo1992} and to ultracold atoms in optical lattices~\cite{Dahan1996}. We also note that the use of the proposed tachometer could be extended to higher angular frequencies while arranging for  $T_{\rm B}$ to still be of the same order of magnitude, but this would place more severe constraints on the control of $y_0$ and $\tau$. 

\subsection{Magnetometer}

Exploiting the formal analogy between inertial and electromagnetic forces implied by Larmor's theorem immediately leads to the possibility of developing also a magnetometer based on the same formal measurement mechanism. As discussed in Refs.~\cite{Coisson1973, Semon1981, Sivardiere1983}, the Hamiltonian of a particle of mass $m$ and charge $Q$ in a magnetic field $\vec{B}$ can be obtained by the substitution $\vec{w}\rightarrow Q\vec{B}/2m$ in Eq. (\ref{Hrot}).

More specifically, the Hamiltonian for a charged particle of mass $m$ and charge $Q$ in an electromagnetic field with vector and scalar potentials $\vec{A}$ and $\phi$ is
\begin{equation}
	H=\frac{(\vec{p}-Q\vec{A})^{2}}{2m}+Q\phi \, ,
\end{equation}
and if the electromagnetic field is a uniform magnetic field $\vec{B}$ along the $z$ direction one has, in the symmetric gauge $\vec{A}=-\vec{r}\times\vec{B}/2$, $\nabla\cdot\vec{A}=0$, 
\begin{equation}
	Q\vec{A}=QB_{z}(x\vec{e}_{y}-y\vec{e}_{x})/2.
\label{eq:qa}
\end{equation}
Considering the same triangular potential and Floquet perturbation as in the previous example, the effective Hamiltonian $H$ for $x$ dimension simplifies then to
\begin{equation}
	H_x=\frac{p_x^2}{2m}+U(x)+V(x)\cos\omega t+\frac{QB_z}{2m}(yp_x-xm\dot{y}),
\label{eq:HX-mag}
\end{equation}
where we have neglected the potential $Q^2B_z^2x^2/8m$, assumed to be much weaker than $U(x)$. This Hamiltonian is formally identical with the Hamiltonian ~(\ref{Hrotx}), with the Floquet-Bloch driving force taking now the explicit form
\begin{equation}
	f=-\frac{Q B_z \eta y_0 \tau }{2\pi m} \, .
\end{equation}
This demonstrates the possibility to exploit this system as a magnetometer, the magnetic field strength is obtained from the Bloch period $T_{\rm B}$ as
\begin{equation}
	B_z=\frac{2\pi m\hbar n}{\eta y_0\tau Q T_{\rm B}}.
\end{equation}

For $B_z\sim 10^{-18}$T, the highest measurement precision obtained by SQUID magnetometers~\cite{Fagaly2006}, a particle with the charge at the single-electron level, $Q\sim 10^{-19}$C, and an atomic scale mass, the resulting weak driving force $f$ results in a long period $T_{\rm B}$ and hence challenging quantum coherence time requirements. To obtain $T_{\rm B}\sim 10^{-3}$s with a perturbation frequency $\omega\sim 10^6$Hz and $\eta\sim 10^{-8}$N, the motion of the particle along the $y$-axis must be controlled at the micron scale, with $y_0\sim 10^{-6}$m and $\tau\sim 10^{-6}$s.

Importantly, we note that the Floquet-Bloch driving force $f$ and the period $T_{\rm B}$ are independent of the electromagnetic gauge, as they should be. For instance in the Landau gauge we have
\begin{equation}
	Q\vec{A}=-QB_z y \vec{e}_x\, ,
\end{equation}
resulting in the probe potential
\begin{equation}
	V_{\rm probe}=\frac{QB_z}{m}yp_x\, ,
\end{equation}
which is different from its form in the symmetric Coulomb gauge, see Eq. (\ref{eq:qa}). However, the factor of 2 difference in the $QB_z/m$ coefficient in the two gauges results in the end in the same value of $f$.

\subsection{Other potentials} 
In addition to the examples considered so far, measurement schemes based on Floquet-Bloch oscillations can be extended to other potentials as well, even to some unusual or singular potentials, by following the same general approach. Examples include the inverse potential $V_{\rm probe}= -a/x$ corresponding to the Coulomb and gravitational forces, the logarithmic potential $V_{\rm probe}= a\log(x)$, and the square-root potential $V_{\rm probe} = a\sqrt{x}$ found in the bound motion of electrons \cite{Gesztesy1978} and quarks \cite{Bose1979, Freitas1978}. Their amplitudes $a$, which are important in many problems in nuclear physics and relativistic quantum mechanics, could in principle be determined by that method. 

In general, however, the action-angle transformation results in a form of the angle variable $\Theta(x,p) $ that is a complicated function of both $x$ and $p$. Still, we show below that an appropriate design of the trapping potential $U(x)$ in $H_{0}$ can result in an approximate angle variable $\Theta(x)$ of the required form to achieve the desired measurement. 

From the equations $m\partial_t^2 x=-\partial_x U(x)$ and $\Theta = \Omega t$ we have that
\begin{equation}
U(x)=-m\int\frac{\partial^2 x}{\partial t^2}dx=-m\Omega^{2}\int \frac{\partial^{2}x}{\partial\Theta^{2}} dx.
\label{Ux}
\end{equation}
Keeping in mind that $\Theta\in\left[0,2\pi\right)$ imposes in general an additional constraint on $x$ to ensure that $U(x)$ is a conservative potential, this equation can be used to find the potential $U(x)$ that results in any desired relationship between $\Theta$ and $x$, 

For the case of the potential $V_{\rm probe}(x) = -a/x$ the determination of $a$ from the Floquet-Bloch period $T_{\rm B}$ requires an action-angle transformation such that $\left |\Theta-\pi\right |=b/x$, where $b$ is some positive constant. From Eq.~(\ref{Ux}) we must then have
\begin{equation}
    U(x)=\begin{cases}
    -\frac{m\Omega_0^{2}x^4}{2b^2} & {\rm for}\; \frac{b}{\pi}<x<L\, ,\\
    \infty & \text{otherwise} \, .
    \end{cases}
\end{equation}
This is a one-dimensional infinite well with a negative quartic bottom profile. When the quartic potential energy dominates the total energy so that $|E_0|\ll m\Omega_0^2 b^2/(2\pi^4)$, this gives, see appendix for details,
\begin{equation}
    a=\frac{\hbar n b}{\tau \Omega_0}\left(\frac{2\pi}{T_{\rm B}}\right)\, .
\end{equation}

\section{Conclusion and outlook}
In this work, we have exploited Floquet engineering techniques to extend the concept of Bloch-oscillations-based precision measurements to the use of Floquet-Bloch oscillations. Specifically, we showed that for a quantum system trapped by some potential $U({\bf r})$, the addition of a weak single-frequency perturbation, combined with a sequence of $\delta$-kicks, results in the onset of the Bloch-like oscillations in its Floquet phase space. These oscillations present significant advantages and flexibility over traditional Bloch oscillations for precision measurements, as a proper design of $U({\bf r})$ and kicking interactions permit to exploit them for the measurement of a broad variety of observables, including rotation rates, magnetic fields, and even the strengths of singular potentials. 

Our proposed scheme requires however the efficient continuous measurement of mean positions or momenta to determine the precise frequency of the Floquet-Bloch oscillations, and the restriction of our analysis to the first quasienergy band implies in addition that this approach is limited to the detection of weak signals.  For signals strong enough to excite interband transitions, a measurement of the emission spectrum of the Floquet system is possible, see e.g. Ref.~\cite{Jiang2022}, and may be more advantageous. A quantitative comparison of the relative benefits and limitations of both approaches will be the subject of future work.

The present paper concentrated on single-particle physics. Floquet-Bloch oscillations are vulnerable to dissipation-induced decoherence, and this imposes as we have seen limitations on the range of accessible Floquet-Bloch frequencies. A possible way to circumvent this problem might involve replacing the system with Floquet time crystals~\cite{Else2016}, which can withstand dissipative environments~\cite{Gong2018, Zhu2019, Hemmerich2021}. We will show in future work how measurement schemes similar to those discussed here can be developed in such systems by using appropriately modulated many-body interactions.

\begin{acknowledgments}
We acknowledge enlightening discussions with Lu Zhou. This work was supported by the Innovation Program for Quantum Science and Technology (Grant No.~2021ZD0303200), the National Key Research and Development Program of China (Grant No.~2016YFA0302001), the National Science Foundation of China (Grants No.~11974116, No.~12234014, and No.~11654005), the Shanghai Municipal Science and Technology Major Project (Grant No.~2019SHZDZX01), and the Fundamental Research Funds for the Central Universities. K. Z. acknowledges the Chinese National Youth Talent Support Program. W. Z. also acknowledges additional support from the Shanghai talent program.
\end{acknowledgments}

\section*{Appendix: Action-angle transformation}
If $H_0$ dominates over all other terms in the Hamiltonian and $H_{1,2}$ can be ignored, the motion of the system is periodic, and the action-angle transform is realized by integration over one period of the motion,
\begin{equation}
	J_0=\frac{1}{2\pi}\oint p dx=\frac{1}{\pi}\int_{x_{\min}} ^{x_{\max}}\sqrt{2m(E_0-U(x))}dx \, , 
    \label{I0}
\end{equation}
where $E_0$ is the energy of the system and $x_{\max, \min}$ represent two peak positions of the periodic motion in the physical coordinates. The angular velocity is given by
\begin{equation}
	\Omega  =\frac{\partial H_{0}}{\partial J_{0}}\,,
	\label{Omega}
\end{equation}
and the relation between $\Theta$ and the physical coordinates is given by the integral
\begin{equation}
	\Theta-\Theta_0=\frac{\partial}{\partial J_0}\int_{x_0}^x p dx \, ,
	\label{theta}
\end{equation}
where $\Theta_0$ and $x_0$ represent their initial values.

For example, if the conservative potential in $H_0$ is a one-dimensional infinite square well,
\begin{equation}
	U(x)=\begin{cases}
		0 \, ,      & 0<x<L            \\
		\infty \, , & \text{otherwise}
	\end{cases}\, ,
\end{equation}
from Eqs.~(\ref{Meff},~\ref{I0},~\ref{Omega}) we obtain
\begin{eqnarray}
	J_0&=&\frac{1}{\pi}\int_0^L\sqrt{2mE_0}=\frac{L}{\pi}\sqrt{2mE_{0}} \, , \label{J0square}\\
	\Omega&=&\frac{\partial E_0}{\partial J_0}=\frac{\pi^2 J_0}{mL^2} = \frac{\pi}{L} \sqrt{2E_0/m}\, , \label{Omegasquare}\\
	M&=&\left(\frac{\partial^2 E_0}{\partial J_0^2}\right)^{-1}=\frac{mL^2}{\pi^2}\, .
\end{eqnarray}
The integral (\ref{theta}) gives
\begin{equation}
	\Theta-\Theta_0=\frac{\partial}{\partial J_0}\int_0^x \sqrt{2mE_0}dx=\frac{\pi}{L}x\,.
\end{equation}
Further taking into account that $x\in \left[0,L\right]$ and $\Theta\in\left[0,2\pi\right)$, we then obtain
\begin{eqnarray}
	x&=&\frac{L}{\pi}\left|\pi-\Theta\right|\, ,\label{xtheta}\\
	p&=&m\dot x=-\sqrt{2mE_0}\ \rm{sign}(\sin\Theta)\, ,
\end{eqnarray}
which also gives
\begin{equation}
	\pi-\Theta={\rm sign}(p)\frac{\pi x}{L}\, .
\end{equation}

Similarly, if the conservative potential is a one-dimensional triangular well
\begin{equation}
	U(x)=\begin{cases}
		\eta x\, , & x\geq 0 \\
		\infty\, , & x<0
	\end{cases}\, ,
	\label{bouncer}
\end{equation}
we obtain
\begin{eqnarray}
	J_{0} & =&\frac{(2E_{0})^{3/2}\sqrt{m}}{3\eta\pi},\\
	\Omega & =&\frac{\eta\pi}{\sqrt{2mE_{0}}}\, ,\label{Omegamgx}\\
	M &=&-\frac{4mE_0^2}{\eta^2\pi^2}\, ,\label{Mmgx}
\end{eqnarray}
and then the transforms,
\begin{eqnarray}
	x &=&\frac{E_0}{\eta}-\frac{\eta(\pi-\Theta)^2}{2m \Omega^2}\, , \label{bouncerx}\\
	p & =&\frac{\eta}{\Omega}(\pi-\Theta)\, .\label{pmgx}
\end{eqnarray}

As a final example, consider the infinite one-dimensional well with a quartic bottom profile 
\begin{equation}
	U(x)=\begin{cases}
		-\frac{m\Omega_0^2 x^4}{2b^2}\, , & \frac{b}{\pi}<x<L \\
		\infty\, ,                        & \text{otherwise}
	\end{cases}\, ,
\end{equation}
with the boundary $L\gg b/\pi>0$. For a small enough initial energy, $|E_0|\ll m\Omega_0^2b^2/(2\pi^4)$, we find the approximate expressions,
\begin{eqnarray}
	J_{0} &=& \frac{\sqrt{2mE_0}}{\pi}\int_{b/\pi}^L\sqrt{1+\frac{m\Omega_0^2x^4}{2b^2 E_0}}\nonumber\\
	&\approx&\frac{m\Omega_0 L^3}{3\pi b}+\frac{E_0}{\Omega_0}-\frac{\pi^4 E_0^2}{10 m b^2 \Omega_0^3}\, ,\\
	\Omega & =&\frac{\partial E_0}{\partial J_0} \approx\Omega_0\, ,\\
	M&=&\left(\frac{\partial^2 E_0}{\partial J_0^2}\right)^{-1}\approx\frac{5mb^2}{\pi^4}-\frac{2E_0}{\Omega_0^2}\, ,
\end{eqnarray}
and then the approximate transforms,
\begin{eqnarray}
	x &\approx&\frac{b}{|\pi-\Theta|+b/L}\, ,\\
	p &\approx&{\rm sign}(\sin\Theta)\sqrt{2mE_0+\frac{m^2\Omega_0 ^2 b^2}{(\Theta-\pi)^4+b^4/L^4}}\, .
\end{eqnarray}

It follows that in the case of a probe potential of the form $V_{\rm probe}(x)= -a/x$, the inverse $x$-dependence results in a linear angular dependence in the phase space,
\begin{equation}
	-\frac{a}{x}\approx \frac{a}{L} -\frac{a}{b}|\pi-\Theta|\, .
\end{equation}

If the probe potential is modulated by a sequence of $\delta$-kicks in alternating directions, 
\begin{equation}
	V_{\rm probe}(x,t)=-\frac{a}{x}{\rm sign}(p)\sum_{l=-\infty}^{+\infty} \delta(t-l T_{\rm D})\,,
\end{equation}
in the Floquet phase space, under the secular approximation (\ref{secapp}), they act as a linear potential $f\vartheta$ with the force,
\begin{equation}
	f=\frac{a\tau}{bT_{\rm D}}.
\end{equation}
So the potential strength $a$ can be obtained from the Bloch period as
\begin{equation}
	a=\frac{\hbar n b T_{\rm D}}{\tau T_{\rm B}}\, .
\end{equation}

\end{document}